\documentclass[onecollarge]{svjour2}
\usepackage{graphicx}
\journalname{Few-Body Systems}
\begin{document}

\title{Analytic expression for three-body recombination rates into deep dimers}
\author{D.V.~Fedorov, M.~Mikkelsen, A.S.~Jensen, N.T.~Zinner}
\institute{IFA, Aarhus University, Aarhus, Denmark}

\date{}
\maketitle

\begin{abstract}

We investigate three-body recombination rates into deep dimers in cold
atomic gases with large scattering length within hyper-spherical adiabatic
zero-range approach.  We derive closed analytic expressions for the rates
for one- and two-species gases.  Although the deep dimers are beyond the
zero-range theory the latter can still describe the recombination into
deep dimers by use of one additional short-range absorption parameter.
The recombination rate, as function of the scattering length, retains the
known universal behavior --- the fourth power trend with characteristic
log-periodic peaks --- however increasing the short-range absorption
broadens the peaks until they are eventually completely smeared out.
Increasing the heavy-to-light mass ratio in a two-species system decreases
the distance between the peaks and increases the overal scale of the
recombination rate.

\end{abstract}

\section{Introduction}

Three-body recombination is the principal source of loss of atoms in
a trapped cold atomic gas.  The loss rates are directly measurable as
functions of the tunable interaction parameters.  A number of recent
experiments have provided results for various combinations of alkali
atoms~\cite{berninger,huang,pires}.

The recombination reaction can conclude with either a shallow ---
weakly bound --- dimer, or with a deep --- strongly bound --- dimer.
The recombination rates into shallow dimers in cold atomic gases
with large scattering length exhibit the well-known universal
behavior -- the rates, as function of the scattering length,
show the fourth power trend with the characteristic log-periodic
structures~\cite{esben-macek,braaten-phys-rep,peder-shallow}. This
behavior is universal -- it can be described by a zero-range theory with
only one parameter, the scattering length.  The universal behavior is
the consequence of the fact that for cold atoms with large scattering
lengths both the three-body dynamics and the shallow dimers depend only on
the long distance properties of the atomic interaction which is largely
determined by the scattering length alone.

The deep dimers, on the contrary, cannot be described by a zero-range
--- a long distance --- theory as their properties depend on the short
distance physics.  For a long distance theory the recombination into deep
dimers, which takes place at short distances, looks simply like a sink
of probability at short distances.  Therefore in zero-range theories
the recombination into deep dimers can be described phenomenologically
by introducing one extra parameter which creates shuch a sink of
probability~\cite{braaten-ann-phys,petrov15,mikkelsen,peder-optical}.

We shall follow this philosophy --- to introduce an extra phenomenological
parameter to account for the loss of probability at short distances ---
and calculate
the recombination rates as function of scattering length
within a hyper-spherical adiabatic zero-range theory.

What distinguishes us from previous investigations is that we derive a
closed analytic expression which works for both one- and two-species
cold gases.

\section{Zero-range model}
\subsection{Hyper-spherical adiabatic approximation}

Within the zero-range hyper-spherical adiabatic approximation~\cite{fedorov93,zrp-reg}
the wave-function of the three-body system is written as
	\begin{equation}
\Psi(\rho,\Omega_{\rho}) =
\rho^{-\frac{5}{2}} f(\rho)\Phi(\rho,\Omega_{\rho}) \,,
	\end{equation}
where the hyper-radius $\rho$ and the hyper-angles $\Omega_\rho$
are defined in appendix~\ref{app:a}.
The hyper-radial function $\Phi(\rho,\Omega_{\rho})$ is normalized to
one in hyper-angles and for large $\rho$ loses its dependence on
$\rho$ and turns into a hyper-spherical harmonic.  Its exact form is of
no importance in the present context.

The hyper-radial function $f(\rho)$ satisfies the hyper-radial equation
	\begin{equation}
\label{eq:hr}
\left(
-\frac{d^2}{d\rho^2}+\frac{\nu^2(\rho)-\frac{1}{4}}{\rho^2}-\frac{2mE}{\hbar^2}
\right)f(\rho)=0 \,,
	\end{equation}
where $m$ is the mass-scale from the definition~(\ref{eq:a:j}) of the
hyper-radial coordinates, $E$ is the energy of the three-body system,
and $\nu(\rho)$ is the angular eigenvalue given by the lowest root of
the angular eigenvalue equation.  The latter depends on the scattering
lengths and masses of the particles.

For a system of three identical particles --- which corresponds to a {\em one-species cold gas} ---
the eigenvalue-equation is given as
	\begin{equation}
\label{eq:a}
-\nu\cos\left(\frac{\nu \pi}{2}\right)+\frac{8}{\sqrt{3}}\sin(\frac{\nu \pi}{6})
 = \frac{1}{\sqrt{\mu_1}}\left(\frac{\rho}{-a_1}\right)\sin\left(\frac{\nu \pi}{2}\right) \,,
	\end{equation}
where $a_1<0$ is the scattering length\footnote{We use the atomic-physics
sign convention where the absence of the shallow dimer corresponds to
negative scattering length.} and $\mu_1$ is dimensionless reduced mass
from the definition~(\ref{eq:a:j}) of the hyper-spherical
coordinates\footnote{For three identical particles we choose $m=m_1=m_2=m_3$ and,
consequently, $\mu_1=\frac{1}{2}$.}.

For a {\em two-species cold gas} we shall consider the common situation
with one light particle with mass $m_3$ and two identical heavy particles
with masses $m_1=m_2$.  We shall assume that only the scattering lengths
between the light and the heavy particles, $a_{1}=a_{2}$, are large and
that the scattering length $a_{3}$ between the heavy particles is,
in comparison, negligibly small.  In this case the angular eigenvalue
equation is given as
	\begin{equation}
\label{eq:a2c}
-\nu\cos\left(\frac{\nu \pi}{2}\right)
-\frac{2}{\sin(2\phi_{12})} \sin\left[\nu\left(\phi_{12}-\frac{\pi}{2}\right)\right]
 = \frac{1}{\sqrt{\mu_1}}\left(\frac{\rho}{-a_1}\right)\sin\left(\frac{\nu \pi}{2}\right) \,,
	\end{equation}
where
	\begin{equation}
\phi_{12}=\arctan\left(\sqrt{\frac{m_3(m_1+m_2+m_3)}{m_1m_2}}\right) \,.
	\end{equation}

\begin{figure}
\centerline{\includegraphics{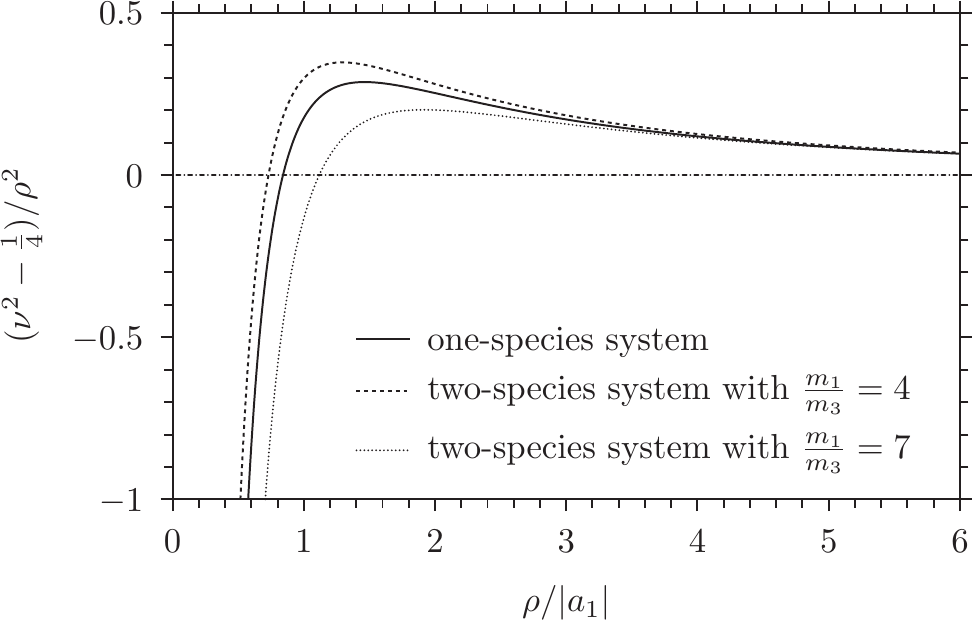}} \caption{The effective potential,
$(\nu^2-\frac14)/\rho^2$, in the hyper-radial equation~(\ref{eq:hr})
for one- and two-species systems.}
	\label{fig:1} \end{figure}

The shape of the effective potential in the hyper-radial
equation~(\ref{eq:hr}) for one- and two-species systems is illustrated on
Fig.~\ref{fig:1}: it features an attractive pocket at short distances
followed by a repulsive barrier at long distances -- an archetypal
shape, conducive to narrow resonances behind a barrier.

\begin{figure}
	\centerline{\includegraphics{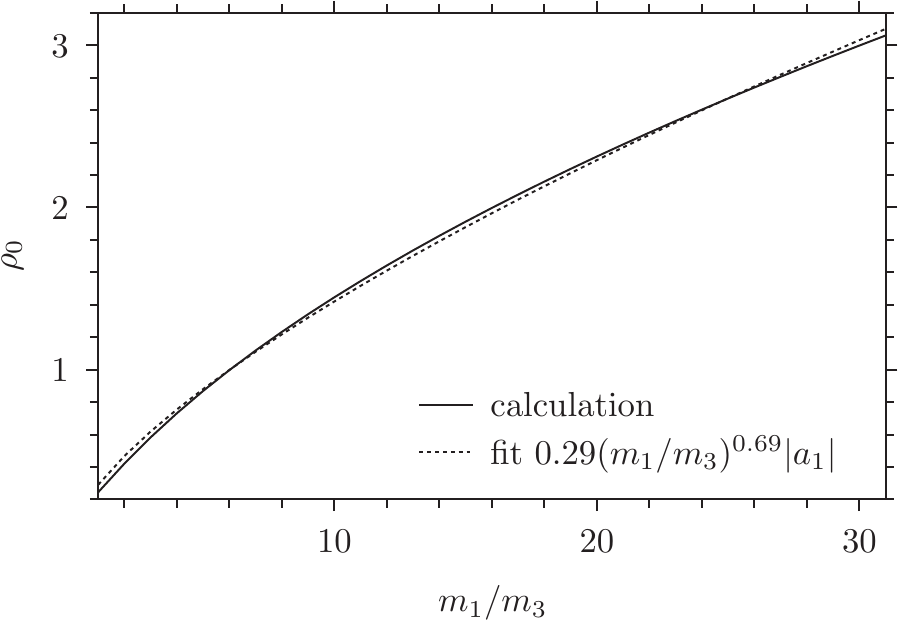}}
	 \caption{
The solution $\rho_0$ to equation $(\nu(\rho_0)^2-\frac14)=0$ for two-species system as function of the heavy-to-light mass ratio
$m_1/m_3$ together with a phenomenological fit.  }
	\label{fig:r0}
	\end{figure}

The dividing
point, $\rho_0$, between the two regions is located where the effective
potential is equal zero, that is, given by the solution to the equation
 $\nu(\rho_0)=\frac12$.
For the one-species system
	\begin{equation}
\rho_0 \approx 0.84 |a_1| \,.
	\end{equation}
For two-species systems we find the following numerical approximation
to $\rho_0$ as function of the heavy-to-light mass ratio $m_1/m_3$
in the practically relevant region of mass ratios,
	\begin{equation}
\frac{\rho_0}{|a_1|} \approx 0.29 \left(\frac{m_1}{m_3}\right)^{0.69} \,.
	\end{equation}
The quality of the approximation is illustrated on Fig.~\ref{fig:r0}.

\subsection{Asymptotic regions}
In the long-distance region, $\rho\gg |a_1|$, the solution to
eigenvalue-equations for both one- and two-species systems is $\nu=2$.  This leads to an
asymptotic equation with a (half-integer) centrifugal barrier,
	\begin{equation}
\left( -\frac{d^2}{d\rho^2}+\frac{4-\frac14}{\rho^2}-\frac{2mE}{\hbar^2}
\right)f(\rho)=0 \,,
	\end{equation}
with the solution
	\begin{equation}\label{eq:asi}
\rho^{-\frac12}f(\rho) = J_{2}(\kappa\rho) - \tan(\delta) Y_2(\kappa\rho) \,,
	\end{equation}
where $J_{2}(\kappa\rho)$, $Y_2(\kappa\rho)$ are the regular and irregular
Bessel functions, $\hbar \kappa=\sqrt{2mE}$, and $\delta$ is the phase-shift
determined by the physics at short distances.

In the short-distance region, $\rho\ll |a_1|$,
the eigenvalue-equation~(\ref{eq:a}) for the one-species system reduces to
	\begin{equation}\label{eq:s}
-\nu\cos\left(\frac{\nu \pi}{2}\right)+
\frac{8}{\sqrt{3}}\sin\left(\frac{\nu \pi}{6}\right)
 = 0 \,,
	\end{equation}
which has an imaginary root $\nu=is$ where $s \approx 1.006$.

\begin{figure}
	\centerline{\includegraphics{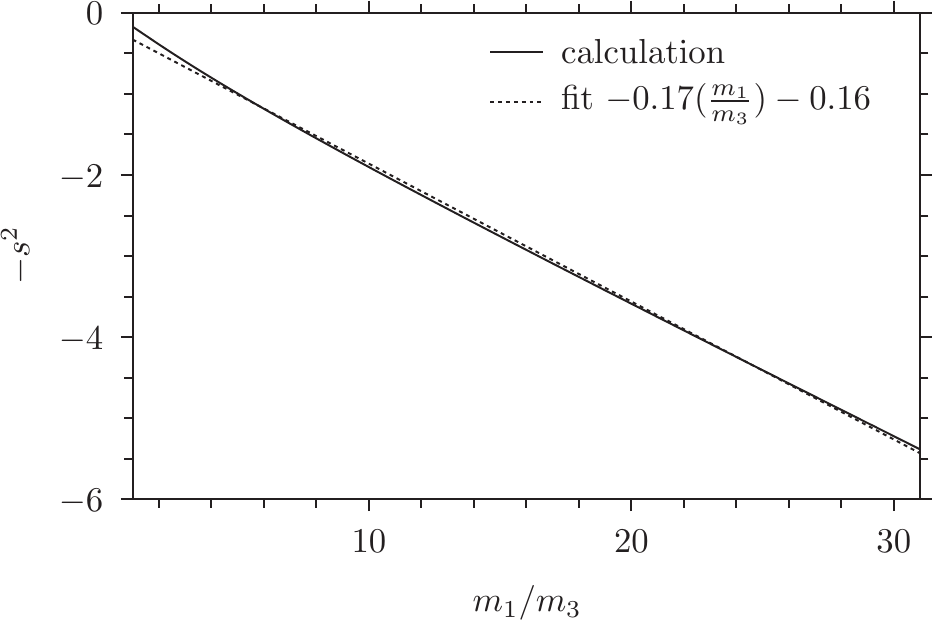}}
	 \caption{
The square, $\nu(0)^2\doteq -s^2$, of the solution to the short-distance eigenvalue
equation~(\ref{eq:s2c}) for two-species system as function of the
heavy-to-light mass ratio $m_1/m_3$, together with a phenomenological
fit.}
	\label{fig:s2c} \end{figure}

The eigenvalue equation~(\ref{eq:a2c}) for two-species system in the short-distance region $\rho\ll |a_1|$ reduces to
	\begin{equation}
\label{eq:s2c}
-\nu\cos\left(\frac{\nu \pi}{2}\right)
-\frac{2}{\sin(2\phi_{12})} \sin\left[\nu\left(\phi_{12}-\frac{\pi}{2}\right)\right]
 = 0 \,,
	\end{equation}
which also has an imaginary root $\nu=is$.  However, the parameter $s$
now depends on the heavy-to-light mass ratio $m_1/m_3$ as illustrated on Fig.~\ref{fig:s2c}: increasing mass ratio
increases $s$, and thus the attraction at short distances.

Thus for both one- and two-species systems the hyper-radial equation at short distances acquires a super-attractive
effective potential,
	\begin{equation}
\left( -\frac{d^2}{d\rho^2}+\frac{-s^2-\frac{1}{4}}{\rho^2}
\right)f(\rho)=0 \,,
	\end{equation}
where we have assumed that the energy is much smaller than the potential.
The equation has two linearly independent solutions,
	\begin{equation}
\sqrt\rho\rho^{\pm is}=\sqrt\rho e^{\pm is\ln\rho} \,,
	\end{equation}
or, equivalently,
	\begin{equation}
\sqrt\rho \sin(s\ln\rho)\,, \; \sqrt\rho \cos(s\ln\rho) \,.
	\end{equation}
The three-body system with zero-range potentials must be
regularized~\cite{peder-shallow} at some short-range scale $r$.  The regularized
wave-function at $\rho\ll |a_1|$ thus takes the form
	\begin{equation}\label{eq:sr}
\rho^{-\frac12}f(\rho) = \sin\left(s\ln\frac{\rho}{r}\right) \,.
	\end{equation}

\subsection{Qualitative solution to hyper-radial equation}
A qualitative solution to the hyper-radial equation can be obtained
analytically by assuming that the two asymptotic solutions --- being
exact correspondingly at $\rho\gg |a_1|$ and $\rho \ll |a_1|$ --- are qualitatively
correct also in the region $\rho \sim |a_1|$. The phase-shift $\delta$
can then be obtained by matching the two asymptotic solutions at $\rho_0$ where attraction turns into repulsion.

At $\rho=\rho_0$
the short-distance function~(\ref{eq:sr}) arrives with the
logarithmic derivative
	\begin{equation}
\left.\frac{\frac{\partial}{\partial\rho}(\rho^{-\frac12}f)}{(\rho^{-\frac12}f)}\right|_{\rho=\rho_0}
= \frac{s}{\rho_0}\cot\left(s\ln\frac{\rho_0}{r}\right)
= \frac{s}{\beta |a_1|}\cot\left(s\ln\frac{\beta |a_1|}{r}\right)
\,,
	\end{equation}
where the factor $\beta=\rho_0/|a_1|$ is given as
	\begin{equation}
\beta=\left\{
\begin{array}{ll}
0.84 & \textrm{, for one-species system} \,; \\
0.29\left(\frac{m_1}{m_3}\right)^{0.69} & \textrm{, for two-species systems} \,.
\end{array}\right.
	\end{equation}
This boundary condition causes the known log-periodic behavior
of certain observables ---
as function of scattering length ---
in three-body
systems with large scattering lengths.

Matching the short-range and the long-range solutions at $\rho=\rho_0$
gives the equation to determine the phase-shift $\delta$,
	\begin{equation}
\kappa
\frac{J_2^\prime(\kappa\rho)-\tan(\delta)Y_2^\prime(\kappa\rho)}
{J_2(\kappa\rho)-\tan(\delta)Y_2(\kappa\rho)}=
\frac{s}{\beta |a_1|}\cot\left(s\ln\frac{\beta |a_1|}{r}\right)
	\,,
	\end{equation}
where the prime denotes the derivative of the Bessel function with respect to its argument.

In the following we shall only be interested in the recombination rate at
zero temperature, that is, at vanishing energies, $\kappa\to 0$.  In this
regime the Bessel functions can be expanded as
	\begin{equation}
J_2(z \to 0) = \frac18 z^2 \,,\; Y_2(z \to 0) = -\frac{4}{\pi}z^{-2} \,,
	\end{equation}
and then the phase-shift in the low-energy limit is given as
	\begin{equation}\label{eq:d}
\tan(\delta) \approx \delta \approx (\kappa \beta |a_1|)^4\frac{\pi}{32}
\frac{1-\frac12 s \cot\left(s\ln\frac{\beta |a_1|}{r}\right)}{1+ \frac12 s \cot\left(s\ln\frac{\beta |a_1|}{r}\right)}
\,.
	\end{equation}

\section{Recombination rate}
\subsection{Reaction rate in terms of hyper-spherical phase-shift}
The long-distance hyper-radial solution~(\ref{eq:asi}) in the limit $\rho\to \infty$ becomes
	\begin{equation}
f(\rho\to\infty) \to \sqrt{\frac{2}{\pi \kappa\rho}}
\frac{e^{-i\kappa\rho+i\phi}+Se^{+i\kappa\rho-i\phi}}{2} \,,
	\end{equation}
where $\phi$ is a real number and $S\doteq e^{2i\delta}$.

If $\delta$ is
a real number then $|S|^2=1$ and the hyper-radial flux density,
	\begin{equation}
j = -i\frac{\hbar}{2m}\left(f^*\frac{\partial f}{\partial\rho}-\frac{\partial f^*}{\partial\rho}f\right) \,,
	\end{equation}
vanishes -- there is no loss of probability if the phase-shift $\delta$ is real.

However, a complex phase shift, $\delta = \Re\delta +
i\Im\delta$, leads to a missing flux density,
	\begin{equation}
\Delta j = \frac{\hbar \kappa}{m}\frac{1-|S|^2}{2\pi \kappa\rho} \,,
	\end{equation}

where
	\begin{equation}\label{eq:1-s2}
1-|S|^2 = 1-e^{-4\Im\delta} \approx 4\Im\delta \,.
	\end{equation}

The total missing flux is obtained by integrating over the hyper-sphere.  The integration element can be obtained from the
relation
	\begin{equation}
\frac{(dV)^2}{d\rho} = \left(\frac{1}{\mu_i\mu_{jk}}\right)^{3/2}\frac{d^3x_id^3y_i}{d\rho}
= m^3\left(\frac{\sum m_i}{\prod m_i}\right)^{3/2}\rho^5 d\Omega_\rho \,,
	\end{equation}
where $dV$ is the three-dimensional volume element and $\int d\Omega_\rho=\pi^3$. Besides this mass factors we also need to collect
the (square of the) factor in front of $J_2(\kappa\rho)$ from the hyper-spherical expansion of
the three-body plane-wave, normalized to unity within three-dimensional volume $V$,
	\begin{equation}
\frac{1}{V}e^{i{\bf k}_x{\bf x}+i{\bf k}_y{\bf y}} =
\frac{1}{V}\frac{(2\pi)^3}{(\kappa \rho)^2}\left(\frac{1}{\pi^3}\right)J_2(\kappa\rho) + \dots \,,
	\end{equation}
where ${\bf k}_x^2+{\bf k}_y^2=\kappa^2$
and ${\bf x}^2+{\bf y}^2=\rho^2$.

Collecting all factors gives the following expression for the probability
loss $\Delta J$ per unit time --- or recombination rate --- within volume $V$,
	\begin{equation}
\Delta J = \frac{1}{V^2} m^3\left(\frac{\sum m_i}{\prod m_i}\right)^{3/2}
2^5 \pi^2 \frac{\hbar}{m} \frac{1-|S|^2}{\kappa^4} \,.
	\end{equation}
The factor $V^2\Delta J$, often referred to as recombination constant $K$,
can be written in terms of the three-body energy $E=\hbar^2\kappa^2/(2m)$ as
	\begin{equation}\label{eq:v2j}
 V^2 \Delta J = m^3\left(\frac{\sum m_i}{\prod
 m_i}\right)^{3/2}8\pi^2\frac{\hbar^5}{m^3}\frac{1-|S|^2}{E^2} \,,
	\end{equation}
directly determines the recombination coefficients in the density balance
equations for cold gases.

\subsection{Recombination rate in zero-range model}
In the zero-range model, where deep dimers are absent, the description
of recombination into deep dimers, unlike recombination
into shallow dimer, requires one extra parameter to make the phase-shift
$\delta$ complex.  This must be a short-range parameter as, indeed,
to recombine into deep dimers all three atoms need to be at
short-distances from each other -- inside the range of their short-range
potentials.  This is also the region where the regularization of the
zero-range three-body model takes place.  Therefore the recombination range
should be close to the regularization range $r$.
The loss of probability --- absorption --- at short distances can be
described by adding an imaginary part to the boundary condition at the
regularization range.  The easiest way to achieve this is to make the
regularization parameter $r$ complex,
	\begin{equation}\label{eq:reps}
r \to re^{-i\epsilon} \,.
	\end{equation}
The absorption parameter $\epsilon$ determines the extent of the
absorption.

Inserting the complex regularization radius~(\ref{eq:reps}) into the
expression for the phase-shift~(\ref{eq:d}), then the phase-shift into
expression~(\ref{eq:1-s2}) for $1-|S|^2$, and the latter, finally, into
the recombination rate~(\ref{eq:v2j}) gives the following expression for
the recombination rate into deep dimers in our complex-boundary-condition
zero-range model,
	\begin{equation}\label{eq:v2je}
V^2\Delta J = m^3\left(\frac{\sum m_i}{\prod m_i}\right)^{3/2} 2^5 \pi^2 \frac{\hbar}{m}
\beta^4 |a_1|^4\frac{\pi}{8}
\Im 
\frac{1-\frac12 s\cot\left(s\ln\frac{\beta |a_1|}{r}+is\epsilon\right)}
{1+\frac12 s\cot\left(s\ln\frac{\beta |a_1|}{r}+is\epsilon\right)} \,.
	\end{equation}
The expression has two phenomenological parameters -- the regularization
range $r$ and the absorption intensity at short distances $\epsilon$.
These parameters cannot be determined within the zero-range model and
should be obtained by fitting the expression to experimental data.

\begin{figure}
\centerline{\includegraphics{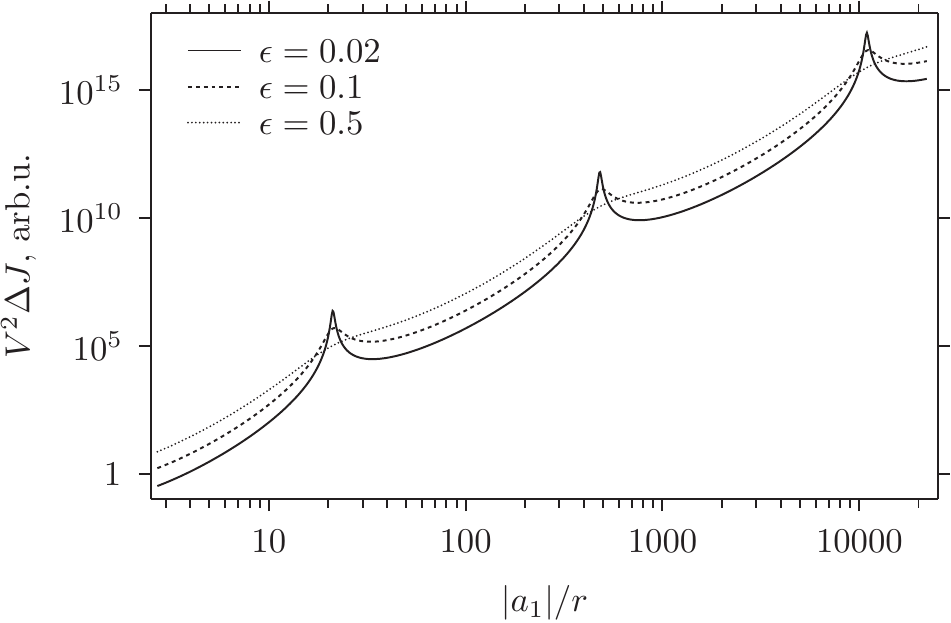}}
\caption{The recombination rate $V^2\Delta J$~(\ref{eq:v2je}) into deep
dimers at zero temperature as function of scattering length $a_1$ for
different values of the absorption parameter~$\epsilon$ for one-species
system.}
	\label{fig:2} \end{figure}

For one-species system
the shapes of recombination rates as function of scattering length
for different absorption parameters are shown on Fig.~\ref{fig:2}.
For small absorption parameter the log-periodic
peaks caused by the term $s\ln(|a_1|/r)$ factor are well formed.
Increasing the absorption smears the peaks out until they eventually disappear
completely.

Two-species systems show similar smearing out of the log-periodic
peaks with increasing the absorption parameter.  However, the shape of the rate is
also influenced by the heavy-to-light mass ratio, as illustrated on
Fig.~\ref{fig:r2c}:
	\begin{figure}
\centerline{\includegraphics{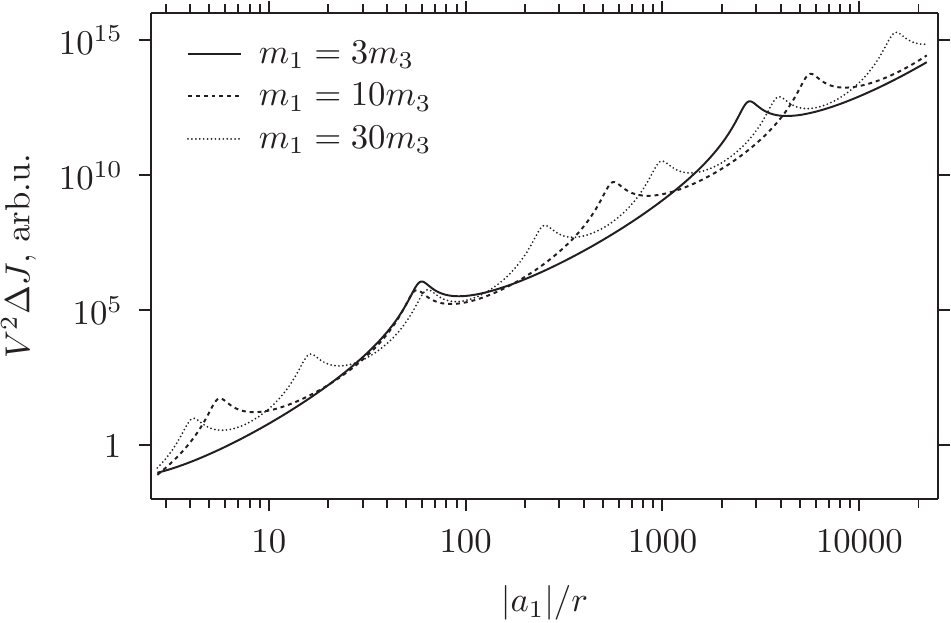}} \caption{The recombination
rate $V^2\Delta J$~(\ref{eq:v2je}) into deep dimers at zero temperature
as function of scattering length $a_1$ for $\epsilon=0.1$ and different
values of the heavy-to-light mass ratio parameter~$m_1/m_3$ for
two-species systems.}
	\label{fig:r2c} \end{figure}
increasing the mass ratio decreases the distance between the peaks and
also increases the overall scale of the rate.

\section{Conclusion}
We have investigated universal properties of the three-body recombination
reaction rates into deep dimers within the
 zero-range hyper-spherical adiabatic approximation. Although
a zero-range model cannot describe the deep dimers, it can still account
for this reaction by introducing one extra parameter, which describes short-range
absorption --- or loss of probability --- in the hyper-radial equation.

We have used a qualitative analytic solution of the hyper-radial equation
comprised of the short- and long-range asymptotic solutions matched at
the middle. This resulted in an analytic expression for the recombination
rate as function of scattering length and the absorption parameter.
We have given the expressions for both one- and two-species cold gases,
the latter with one light and two identical heavy particles.

For weak absorption the recombination rate as function of scattering
length shows characteristic log-periodic peaks. Increasing the absorption
broadens the peaks until they are eventually smeared out.

For a meaningful comparison with experimental data the effects
from the scattering length between the heavy particles need to be
included~\cite{mikkelsen}.  The third scattering length modifies both the
$s$ and $\beta$ parameters in the expression for the recombination rate
in a non-trivial way through significant modification of the eigenvalue
equaition.  This should be a subject of a separate investigation.

\appendix
\section{Hyper-spherical coordinates}
\label{app:a}
Given the particle
coordinates $\mathbf{r}_i$ and masses $m_i$, where $i\in\{1,2,3\}$,
the Jacobi coordinates are defined as
	\begin{equation} \label{eq:a:j}
\mathbf{x}_i=\sqrt{\mu_i}(\mathbf{r}_j-\mathbf{r}_k) \,, \;
\mathbf{y}_i=\sqrt{\mu_{jk}}\left(\mathbf{r}_i-\frac{m_j
\mathbf{r}_j+m_k\mathbf{r}_k}{m_j+m_k}\right)
\label{eq:jaccord}
	\end{equation}
where 
	\begin{equation}
\label{eq:reducedmass}
\mu_i=\frac{1}{m}\frac{m_j m_k}{m_j+m_k} \,, \; \mu_{jk}=\frac{1}{m}\frac{m_i(m_j+m_k)}{m_i+m_j+m_k} \,.
	\end{equation}
where $m$ is an arbitrary mass-scale.

The hyper-angular coordinates are the hyper-radius, $\rho$, defined as
	\begin{equation}
\label{eq:hypersphericalcoord}
\rho^2 \doteq \mathbf{x}_i^2+\mathbf{y}_i^2  = 
\frac{\sum_{i<k}m_i m_k (\mathbf{r}_i - \mathbf{r}_k)^2}
{m(m_1+m_2+m_3)} \; ,
	\end{equation}
and five hyper-angles, $\Omega_\rho$,
	\begin{equation}
\Omega_\rho \doteq \left\{\alpha_i\doteq\arctan\left(\frac{x_i}{y_i}\right),\frac{\mathbf{x}_i}{x_i},\frac{\mathbf{x}_i}{x_i}\right\} \,.
	\end{equation}

For identical particles the index can be omitted.

\end{document}